\documentclass[conference]{IEEEtran}
\IEEEoverridecommandlockouts
\usepackage{cite}
\usepackage{amsmath,amssymb,amsfonts}
\usepackage{algorithmic}
\usepackage{graphicx}
\usepackage{amssymb}
\usepackage{textcomp}
\usepackage{hyperref}
\usepackage{xcolor}
\usepackage{flushend}

\def\BibTeX{{\rm B\kern-.05em{\sc i\kern-.025em b}\kern-.08em
    T\kern-.1667em\lower.7ex\hbox{E}\kern-.125emX}}
\begin{document}

\title{MaMIMO CSI-based positioning using CNNs: \\ Peeking inside the black box}

\author{\IEEEauthorblockN{Sibren De Bast, Sofie Pollin}
\IEEEauthorblockA{\textit{ESAT - TELEMIC} \\
\textit{KU Leuven}, Belgium\\
sibren.debast@kuleuven.be}
}

\maketitle

\begin{abstract}
Massive MIMO (MaMIMO) Channel State Information (CSI) based user positioning systems using Convolutional Neural Networks (CNNs) show great potential, reaching a very high accuracy without introducing any overhead in the MaMIMO communication system. In this study, we show that both these systems can position indoor users in both Line-of-Sight and in non-Line-of-Sight conditions with an accuracy of around 20 mm. However, to further develop these positioning systems, more insight in how the CNN infers the position is needed. The used CNNs are a black box and we can only guess how they position the users. Therefore, the second focus of this paper is on opening the black box using several experiments. We explore the current limitations and promises using the open dataset gathered on a real-life 64-antenna MaMIMO testbed. In this way, extra insight in the system is gathered, guiding research on MaMIMO CSI-based positioning systems using CNNs in the right direction. 
\end{abstract}


\section{Introduction}
In recent years, Massive Multiple-Input Multiple-Output (MaMIMO) communication systems have revolutionised the way cellular networks use the physical layer to transfer data to user terminals \cite{mamimo}. It does so by deploying tens to hundreds of antennas at the Base Station (BS) and modulating the signals in such a way that the system is able to serve multiple users in the same time and frequency slot. This is done by multiplexing the users in the spatial domain using beamforming methods. In order to achieve spatial focusing, the Channel State Information (CSI) needs to be known for every user.

This Channel State Information is collected by the use of an uplink pilot symbol. All users send a known symbol during a predefined timeslot. The BS uses this signal to estimate the Channel Response of the channels between all users and all BS-antennas. Once this is done, the BS can use the CSI to decode and precode the up-link and down-link data. Since the system is already required to estimate the CSI in order to operate the communication system, it introduces no extra overhead in the communication system to use the CSI to estimate the location of the user.

E. Björnson et al. envision that in the near future MaMIMO system will enable accurate six-dimensional positioning of the user terminals \cite{bjornson}. These six dimensions include next to the spatial location (three dimensions) also the orientation (pitch, yaw, and roll) of the users terminal. However, the authors also identify a need for open, very precisely spatially labelled datasets, consisting out of MaMIMO CSI and the according position of the user. These are needed to develop new MaMIMO models that can be used to study the positioning capabilities of MaMIMO systems.

The use of Machine Learning (ML) techniques, in particular Convolutional Neural Networks (CNNs), to extract the spatial information from MaMIMO CSI, was introduced by Vieira et al. \cite{lund}. They showed the feasibility of the idea using CSI-samples generated by simulating an environment with the one-ring channel model (COST 2100 MIMO model \cite{cost2100}). Furthermore, the presented CNN model reached a high positioning accuracy. However, they learned the statistics of a channel model using a CNN model. Therefore, this combination of ML and simulated data always needs to be evaluated using real data before real conclusions could be drawn. Yet recently, some research using measured datasets has surfaced. 

Widmaier et al. generated a spatially labelled dataset of a large corridor in their office building. They showed a fair localisation accuracy of the CNN based positioning solution. \cite{stuttgart} But, this accuracy was lower than the accuracy achieved by Vieira et al. \cite{lund}. Furthermore, the authors presented some experiments to test the robustness and reproducibility over time of the proposed system. They found that the systems accuracy was affected by moving pedestrians and a changing propagation environment (closing and opening of windows and doors). 

To further explore the limits to the accuracy of the proposed MaMIMO CSI-based localisation systems using CNNs, we generated three indoor spatially labelled dense open datasets \cite{sibren}. Here, each dataset represented the same room but the MIMO array topology was changed. All of these measurements were done in a static Line-of-Sight (LoS) propagation environment. In this earlier work, we showed a positioning accuracy of 55.35 mm (0.48 $\lambda$). This was achieved through the high density and size of the datasets, consisting of 252004 CSI-samples each covering an area of around 3 by 3 metres. Furthermore, we showed that transfer learning can lower the need for such a large dataset when transferring knowledge from a model trained on one array topology to a model that has to be trained for another topology. As a result, the amount of samples can be decreased to 5000 while still achieving the same level of accuracy. When combining transfer learning and 100000 new data points we could even improve the localisation accuracy to a mean error of 23.92 mm (0.208 $\lambda$). This paper will explore a new CNN architecture to further improve upon this accuracy.

These previous results have shown that the use of CNNs to extract spatial information from CSI is a very efficient and promising method to localise users in a MaMIMO system. Nonetheless, they do not provide ample insights into how these models use the CSI and how we can increase their performance in more challenging conditions e.g.: when people are present and non-Line-of-Sight (nLoS) scenarios.

This work will focus on gathering useful information on how these systems work. Which components of the CSI does the CNN use to position the users? What is the influence of movement in the proximity of the users and does all movements influence the system in equal amounts? We will try to peek inside the black box of the CNNs. To fulfil this target, two experiments are designed.

The structure of the paper is as follows:
First, Section \ref{sec:cnn} defines the architecture of the proposed CNN. Next, we discuss the first proposed experiment in Section \ref{sec:experimentI}. The second experiment is processed in Section \ref{sec:experimentII}. To end, we conclude the paper in Section \ref{sec:conclusion}.

\section{Convolutional Neural Network Model}
\label{sec:cnn}
In this section, first, the implementation of the Convolutional Neural Network (CNN) used for this task is delineated. Afterwards, the performance on previously published and new datasets is explored.

\subsection{Convolutional Neural Network Architecture}
CNNs are a type of Neural networks that apply convolutions on the data with trainable filters to extract features and information. They have proven to be very efficient in extracting structured features from complex data, for example, they have revolutionised the world of computer vision by taking a big accuracy leap in classifying images depending on their content. 

During the training of these CNNs, the filters used during the convolutions are altered using back-propagation and tuned to perform a specific task. In this case, the task is accurately positioning users of a MaMIMO communication system using the provided CSI. The CNN is used to extract relevant features from a user's CSI and convert is to a lower dimensional vector, which can be used to determine the position of the associated user.

To complete the task and position the users based on their CSI, the CNN will have to learn how to extract spatial information out of the data. These spatial features are quite complex, therefore, a deep CNN has to be designed to achieve the goal. The deeper CNNs are made, the harder it becomes to update the trainable parameters of the higher layers (layers in the beginning of the network) during the training process. This is due to the gradient that has to pass through many layers and will start to vanish before reaching these higher layers. This problem is referred to as the vanishing gradient problem. In recent years, many CNN architecture improvements are proposed to counteract this problem, many of which use a type of skip-connections to help propagate the gradient. In this work, we use an architecture based on DenseNet\cite{densenet} to design the CNN.

\begin{figure}[!ht]
\centering
\includegraphics[width=0.8\linewidth]{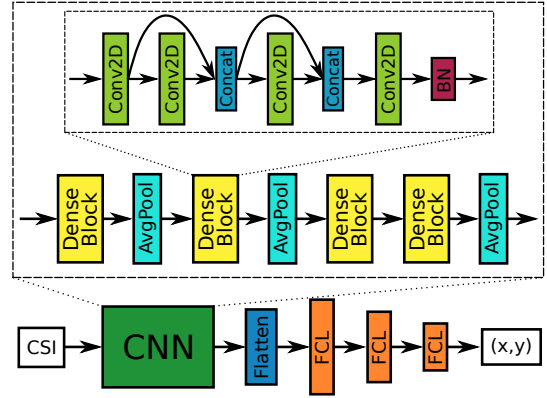}
\caption{The used architecture for the CNN. It is based on the DenseNet architecture and consists out of 16 Convolutional layers and 3 Fully Connected Layers (FCL). The full details can be found in the source code available at GitHub.}
\label{fig:cnn}
\end{figure}

Figure \ref{fig:cnn} shows the architecture of the CNN used in this paper. It consists of two main parts, a CNN-based part followed by three Fully Connected Layers (FCL). The CNN is constructed using "Dense Blocks". These Dense Blocks contain four convolutional layers where the size of the feature maps remains equal and the data is propagated both through convolutional layers as well as through skip-connections based on concatenation layers. At the end of each Dense Block, a Batch Normalisation (BN) layer is added to reduce over-fitting. In between Dense Blocks, an Average Pooling layer reduces the size of the feature maps to reduce the dimensions of the data. At the end, the FCLs combine the features extracted by the CNN into the position of the user.

This deep learning model was implemented in the Keras ML framework using the TensorFlow back-end. 
The source code of the model is published on GitHub\footnote{\url{https://github.com/sibrendebast/inside-the-black-box}} together with all other scripts used to generate the results of this paper.

\subsection{CSI-based Positioning Performance}

To evaluate the performance of the proposed CNN architecture, we trained the CNN on three different datasets and evaluated its ability to accurately position the users. All datasets are recorded using the same 64-antenna MaMIMO base station, deployed as an 8 by 8 Uniform Rectangular Array (URA). The system uses a centre frequency $f_c$ of 2.61 GHz ($\lambda$ = 114.56 mm) and a bandwidth of 40 MHz.

\begin{figure}[!ht]
    \centering
    \includegraphics[width=0.9\linewidth]{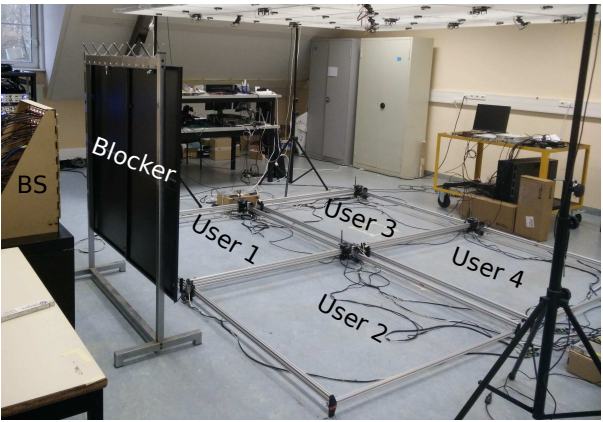}
    \caption{The measurement environment used to create the lab LoS and nLoS datasets. For the creation of the LoS dataset, the metal blocker was removed.}
    \label{fig:lab}
\end{figure}

The first dataset was recorded in the boardroom of our office building with all users in direct LoS of the base station. The full details of the creation of this dataset is presented in \cite{sibren}. This room was chosen for its simple and clean layout, to minimise the scattering of the signals in the environment. The second and third dataset are new datasets, created in the same way as the first dataset. They were recorded in our MIMO-lab, where the rich scattering environment will strongly influence the nLoS components in the CSI. A picture of this measurement set-up can be seen in Figure \ref{fig:lab}. For the second dataset, the users were in LoS of the base station, while for the third dataset, a metal blocker was placed in between the base station and the user to block the direct path. Each dataset consists out of 252004 samples and are publicly available\footnote{\url{https://homes.esat.kuleuven.be/~sdebast}}.

These CSI samples consist out of one channel measurement from one user. For each user, 64 antennas record an IQ sample for a total of 100 subcarriers. This gives rise to a complex channel tensor $\mathbf{H} \in \mathbb{C}^{64 \times 100}$. Since neural network can not process complex numbers (for now), the real and imaginary parts are split and we end up with a thrid dimension, resulting in a channel tensor of the form $\mathbf{H} \in \mathbb{C}^{64 \times 100 \times 2}$. This is used by the CNN as input.


The proposed CNN was trained three times, one time on each of these datasets. The training set consisted of 85\% of the available samples. To fine-tune the training, a validation set with a size of 5\% of the available data was used. Finally, the performance was tested using a test set containing the remaining 10\% of the dataset. The positioning accuracy of the proposed CNN on these three datasets can be seen in Table \ref{tab:perf}.

As can be seen, the model reaches an accuracy of 17.16 - 17.30 mm (0.150 - 0.151 $\lambda$) on the LoS datasets and 20.26 mm (0.176 $\lambda$) on the nLoS dataset. For the boardroom dataset, this new model more than doubles the accuracy in comparison to the previous reported mean estimation error of 55.35 mm (0.48 $\lambda$)\cite{sibren}. This is due to the architecture of the proposed model. This model has many more Convolutional layers in comparison to the model proposed in \cite{sibren} and is therefore a deeper model. In general, a deeper model can learn more complex features than a more shallow model. However, this comes at the cost of slower training and the vanishing gradient problem. During the design of the new proposed model, these two problems were taken into account, resulting in an architecture that is able to train the higher layers effectivly in an efficient way. As a result, the deeper model is able to improve on the positioning accuracy of the state-of-the-art.

\begin{table}[!ht]
\centering
\caption{Mean error on the positioning datasets.}
\begin{tabular}{|r|r|r|}
\hline
Boardroom, LoS  & MIMO-lab, LoS & MIMO-lab, nLoS \\
\hline
17.16 mm        & 17.30 mm      & 20.26 mm \\
0.150 $\lambda$ & 0.151 $\lambda$  & 0.176 $\lambda$ \\
\hline
\end{tabular}
\label{tab:perf}
    
\end{table}

\section{Experiment I: Influence of the direct path.}
\label{sec:experimentI}
\begin{figure}[!ht]
    \centering
    \includegraphics[width=0.7\linewidth]{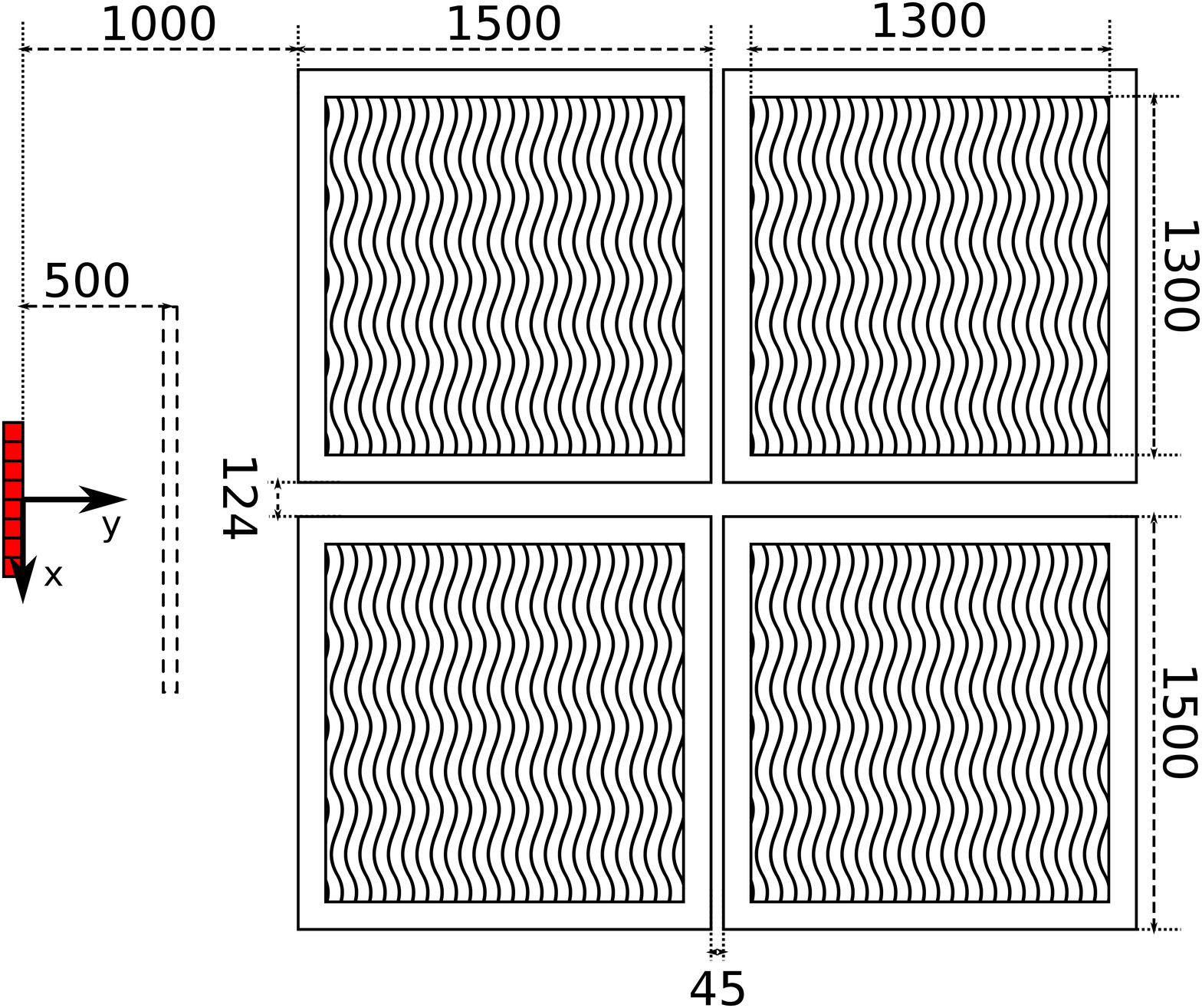}
    \caption{The scenario used in the first experiment.}
    \label{fig:exp1}
\end{figure}

In the first experiment, the effect of the LoS components is explored.
We consider two open datasets that we created. Both deploy the MaMIMO BS using the same array topology and the users are placed in exactly the same region in both scenarios. However, the system was deployed in two different rooms, the boardroom and the MIMO-lab. Therefore, the direct path between the BS and the users should be relatively equal, but the multi-path components should be totally different. Figure \ref{fig:exp1} gives an overview of how how the users are located in the rooms. An 8-by-8 URA is deployed in front of the room, this is depicted by the red squares. The users are positioned by four XY-tables, moving the users accurately inside the wavey regions. This resulted in two datasets containing measured CSI-samples with the same direct-path components, but completely different multi-path components.

If the system only uses the direct path to position the users in the space, the model would still be able to roughly estimate the position of the users in the new environment, since the direct path is not altered in a significant way. When this is not the case, the model also needs the multi-path components, or a combination of both, in order to effectively extract the spatial information from the CSI.

The model was trained on the LoS dataset recorder in the boardroom. Next, the CSI-samples of the test set of the LoS dataset recorded in the MIMO-lab were used by this trained model to estimate the position of the users. On this test set, the model gained an accuracy of 701.76 mm. This result is only slightly better than randomly guessing a location inside the considered area of possible positions. 

To verify that the poor result was not due to a misalignment during the set-up of the scenario, the mean direction of the estimation error is calculated. If the poor results is due to a mismatch in placement of the BS in relation to the users, between the measurements in the boardroom and the MIMO-lab, this mean direction will have a magnitude in the order of the size of the mean estimation error. However, the mean direction of the estimation error was 11.59 mm in the x-direction and -21.59 mm in the y-direction. 

This experiment shows that CSI-based localisation of user in MaMIMO systems using CNNs in LoS scenarios, does not only rely on the direct path to position the user. Moreover, it needs to learn the multi-path components of the channels, which heavily depends 
on the topology of the room, in order to accurately estimate the position of the users.

\section{Experiment II: influence of nomadic environments.}
\label{sec:experimentII}

In this experiment, the influence of movement in the proximity of the users is assessed. The localisation is based in the CSI of a particular user at a particular time. This CSI contains a signature of all paths the signal can use to travel between the antenna of the user and the many antennas of the base station. When an object is moving in the proximity of the user, these paths will be altered. This experiment wants to examine how much the positioning accuracy of the proposed CNN is affected when objects are moving in the proximity of the users, while the model is only trained for data in a static environment. How much will the estimated position deviate from the actual position? Which movements will affect the localisation the most? How can we make the localisation system more robust?

In order to answer these questions, several measurements were performed. All of these measurements have been performed using the same method: The four user-positioners were moved to the centre of their area. The positioners were kept static while a person - me, the first author - was walking back and forth along a defined path during two minutes. During this period, the base station collected the CSI from the four users every 0.5 s, resulting in 240 samples per user. This experiment was repeated seven times, giving 6 nomadic scenarios and one reference scenario:

\begin{enumerate}
    \item Moving at the back;
    \item Moving at the left side;
    \item Moving at the right side;
    \item Moving in the front;
    \item Moving in the middle left to right;
    \item Moving in the middle front to back;
    \item No movement (reference).
\end{enumerate}

\begin{figure}[!ht]
    \centering   
    \includegraphics[width=0.7\linewidth]{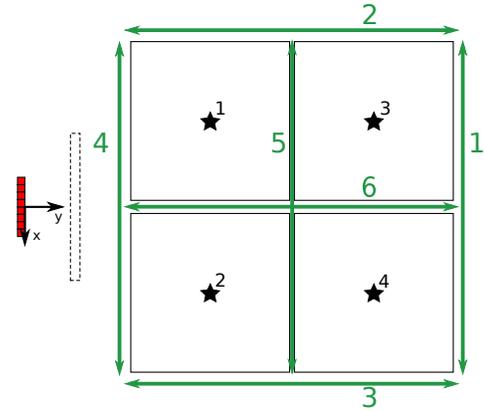}
    \caption{The scenario of the second experiment. The four users are positioned in the middle of the reachable area. To test the influence of moving objects on the positional accuracy, I - the first author - walked back and forth along six different trajectories while taking CSI measurements. These trajectories are depicted using green arrows. The striped lines represent the blocker to create the nLoS scenarios.}
    \label{fig:exp2}
\end{figure}

The defined paths from the nomadic scenarios are shown in Figure \ref{fig:exp2}.

\subsection{Results}

Using these measurements, the influence of movement in the proximity of the users can be checked by means of assessing how much the estimated position of the user changes due to changing environment. This is done by calculating the deviation of the estimated position. To do so, two neural networks, implemented as presented in Section \ref{sec:cnn} were trained, one on the MIMO-lab LoS dataset and one on the MIMO-lab nLoS dataset. Both of them are tasked to estimate the location of the user with corresponding channels, measured during Experiment II.

Under the influence of the movement in the environment, the measured channel will vary, therefore, the estimated position will change over time. To check how much the estimation deviates, the measurements from the static (no movement) scenario are used as a reference. The estimated positions of this control group are averaged and this mean is used as reference. Next, all other CSI samples are used to estimate the position and the deviation of this estimated position is calculated using the references.

First of all, we examine the mean deviation of the estimated position. The mean deviation is calculated for each movement scenario and for each of the four users. For the LoS case, the results are shown in Table \ref{tab:mean_deviation}. 
When a user's direct path is blocked during the measurements due to the movements, the associated mean deviation is printed in bold.

\begin{table}[!ht]
\centering
\caption{The mean deviation [millimetre] recorded in experiment II for each user in each nomadic scenario for the LoS case. 
}
\begin{tabular}{|l||r|r|r|r|r|r|r|}
\hline
User & Ref      & 1         & 2         & 3         & 4                 & 5                 & 6\\
\hline
1    & 5.61     & 42.48     & 63.31     & 39.48     & \textbf{268.50}   & 63.98             & 94.10 \\
2    & 8.80     & 59.85     & 49.83     & 74.36     & \textbf{285.42}   & 157.84            & 127.47 \\
3    & 21.17    & 69.05     & 99.73     & 196.08    & \textbf{183.18}   & \textbf{187.97}   & \textbf{328.56} \\
4    & 8.69     & 104.88    & 64.10     & 86.32     & \textbf{240.80}   & \textbf{277.41}   & \textbf{310.54} \\
\hline
\end{tabular}
\label{tab:mean_deviation}
\end{table}

The first thing to notice is, when the environment is static, the deviation of the estimated position is very stable. Once some movement is added to the environment, the deviation begins to grow. The type of movement can be split into two groups, movements that block the direct path and those that do not. When a movement is not blocking the LoS, the accuracy of the system lowers slightly, with a factor of 2 to 5 times, but is still able to position the users within 10 cm. This can be observed in scenario 1, where the moving person was behind all users in relation to the BS. However, when the direct path is blocked, the results printed in bold, the accuracy decreases substantially. This can be observed in scenario 4, 5, and 6.

For the nLoS case, the mean deviation is shown in Table \ref{tab:mean_deviation_nlos}. In comparison to the LoS scenarios, nLoS based positioning is more robust to movement in the environment. This can be attributed to the many components in the multi-path channel in the CSI that can be used to localise the user. The model has learned to use many more paths to localise the user and therefore, has gained some redundancy that it can use to position a user correctly when some of the paths are altered. 
\begin{table}[!ht]
\centering
\caption{The mean deviation [millimetre] recorded in experiment 2 for each user in each scenario in the nLoS case. 
}
\begin{tabular}{|l||r|r|r|r|r|r|r|}
\hline
User & Ref      & 1         & 2         & 3         & 4        & 5        & 6\\
\hline
1    & 15.23    & 49.94     & 126.40    & 83.63     & 722.03   & 85.24    & 131.21 \\
2    & 15.83    & 35.29     & 37.91     & 47.52     & 77.23    & 98.51    & 61.86 \\
3    & 17.96    & 48.85     & 47.15     & 40.33     & 204.56   & 136.93   & 54.23 \\
4    & 33.61    & 45.97     & 43.94     & 66.42     & 162.05   & 94.54    & 87.68 \\
\hline
\end{tabular}
\label{tab:mean_deviation_nlos}
\end{table}


\begin{figure}[!ht]
\centering
\includegraphics[width=\linewidth]{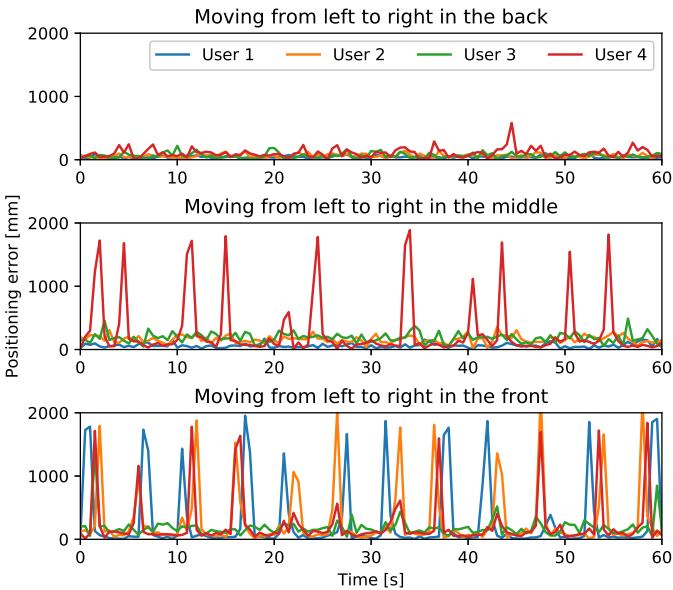}
\caption{The measured localisation deviation during the first minute of the LoS measurements of scenario 1, 5, and 4. The accuracy diminishes substantially when the direct path is blocked.}
\label{fig:dyn_los}
\end{figure}

\begin{figure}[!ht]
\centering
\includegraphics[width=\linewidth]{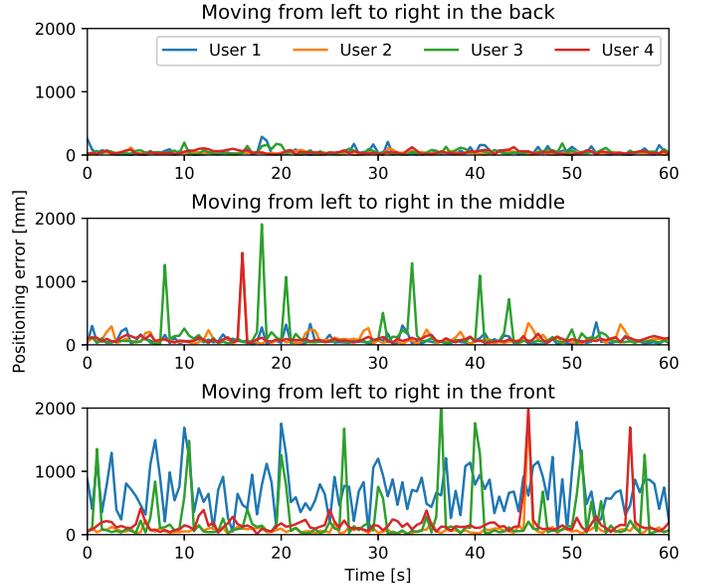}
\caption{The measured localisation deviation during the first minute of the nLoS measurements of scenario 1, 5, and 4. The movement affects the accuracy a lot less in comparison to the LoS case as seen in Fig. \ref{fig:dyn_los}. However, it becomes a lot less predictable when the accuracy will be affected significantly.}
\label{fig:dyn_nLoS}
\end{figure}

Next, the deviation over time due to the movement is evaluated. Figure \ref{fig:dyn_los} and Figure \ref{fig:dyn_nLoS} show how the deviation is affected over due to the nomadic environment. The figures show the positioning deviation during the first minute of the measurements of scenario 1, 5, and 4. In the LoS case, the deviation is particularly large when the direct path between the user and the BS is blocked, this gives rise to a periodic spike in the deviation of the estimated error. The size of the estimation error reached up to 2 m, which is particularly poor when regarding the total size of the studied area. 

When assessing the deviation over time of the estimated position in the nLoS scenarios, the model again shows an increased robustness against a nomadic environment. In many cases, the accuracy is only affected in a minor way. However, in some cases, for example user 1 in scenario 4, the positioning accuracy is affected all the time, no matter where the moving person was located in the room. Furthermore, the periodic behaviour of the deviation is missing in the nLoS case. This results in a lower predictability of when the estimated position will be accurate or not. This in contrast to the LoS case, here we could notice that the position was accurate when the user was in direct LoS of the BS.

\section{Peeking inside the back box: Conclusions}
\label{sec:conclusion}

In this paper, we presented a Convolutional Neural Network based model, able to accurately position MaMIMO users, by employing the Channel State Information. The model reached for both Line-of-Sight and non-Line-of-Sight scenarios a high accuracy of 17.16 (0.150 $\lambda$) to 20.26 mm (0.176 $\lambda$). This improves the current state-of-the-arts accuracy more than 50\%. We explored how the model was able to reach this performance and found that it needs both the direct and non-direct paths between the users and base station. The model learns the position of the users by learning the complex combination of these components. 

When the model is deployed in a scenario where the direct paths are identical, but the environment is completely different, the model is unable to accurately position the the users. This tells us that the model needs the multi-path components in the signal to perform its task. 

When the model is estimating the position of users in a nomadic environment, in this case a person was moving in the room, the accuracy is affected. This is due to the changing multi-path components in the CSI. The extent of this influence depends if the user has a direct path component in its channel. In the LoS case, the ability to locate the users in the room suffers somewhat due to the changing multi-path. But only when the direct path is blocked, the model is unable to position the user in the room. Therefore, it is quite predictable how much the positioning accuracy will be affected in a primarily LoS scenario. 

On the contrary, when a user has a nLoS connection with the BS, the model is more robust to nomadic environments by exploiting the many multi-path components and the diversity they bring, to position the users. This leads to a more reliable positioning system where users can still be positioned to a certain extent while the environment changes slightly. However, sometimes, the environment still changes too much to localise the users. Furthermore, it is unpredictable when this is the case.

\section{Future work}

The conclusions of this work can be used to guide future research. First of all, in LoS scenarios, the direct path to the antenna array should not be blocked in order for the current models to work. This loss of LoS can be averted using a distributed antenna array, since the probability that the LoS link with the majority of antennas is broken is much smaller in this case. Therefore, distributed MaMIMO systems should be studied and evaluated if they do in fact improve the robustness of the positioning system in LoS scenarios. Furthermore, distributed MaMIMO systems can also improve the diversity of the multi-path, targeting a higher robustness in nLoS positioning as well.

\section{Acknowledgement}
This research was partially funded by the Research Foundation Flanders (FWO) SB PhD fellowship, grant no. 1SA1619N (Sibren De Bast).

\end{document}